\documentclass{llncs}
\usepackage{amsmath}
\usepackage{booktabs} 
\usepackage{caption} 
\usepackage{graphicx}
\usepackage{pgfplots}
\usepackage[all]{nowidow}
\usepackage[utf8]{inputenc}
\usepackage{tikz}
\usetikzlibrary{er,positioning,bayesnet}
\usepackage{multicol}
\usepackage{hyperref}
\usepackage[inline]{enumitem} 
\usepackage{wrapfig}
\usepackage[caption=false]{subfig}

\definecolor{blue}{HTML}{1F77B4}
\definecolor{orange}{HTML}{FF7F0E}
\definecolor{green}{HTML}{2CA02C}

\pgfplotsset{compat=1.14}

\usepackage{multirow}
\usepackage{amsmath,amsfonts,mathtools}
\usepackage{diagbox}
\usepackage{amssymb}

\usepackage{algorithm}
\usepackage[noend]{algpseudocode}

\algnewcommand\algorithmicforeach{\textbf{for each}}
\algdef{S}[FOR]{ForEach}[1]{\algorithmicforeach\ #1\ \algorithmicdo}

\usepackage[font=footnotesize]{caption}
\usepackage{subfig}
\usepackage{booktabs} 
\usepackage{multirow}
\usepackage{amsmath,amsfonts,mathtools}
\usepackage{diagbox}

\algtext*{EndWhile}
\algtext*{EndIf}
\algtext*{EndFor}

\usepackage{pifont}
\newcommand{\xmark}{\ding{55}}%

\begin{document}

\title{\emph{ExTru}: A Lightweight, Fast, and Secure \emph{Ex}pirable \emph{Tru}st for the Internet of Things}

\author{Hadi Mardani Kamali \and Kimia Zamiri Azar \and \\ Shervin Roshanisefat \and Ashkan Vakil \and Avesta Sasan}

\institute{Department of Electrical and Computer Engineering, \\
George Mason University, Fairfax, VA, USA. \\
\email{\{hmardani, kzamiria, sroshani, avakil, asasan\}@gmu.edu}}

\maketitle

\begin{abstract}

The resource-constrained nature of the Internet of Things (IoT) devices, poses a challenge in designing a secure, reliable, and particularly high-performance communication for this family of devices. Although side-channel resistant ciphers (either block cipher or stream cipher) are the well-suited solution to establish a guaranteed secure communication, the \emph{energy}-intensive nature of these ciphers makes them undesirable for particularly lightweight IoT solutions. In this paper, we introduce \emph{ExTru}, a novel encrypted communication protocol based on stream ciphers that adds a configurable switching \& toggling network (\emph{CSTN}) to not only boost the performance of the communication in lightweight IoT devices, it also consumes far less energy compared with the conventional side-channel resistant ciphers. Although the overall structure of the proposed scheme is leaky against physical attacks, such as side-channel or new scan-based Boolean satisfiability (\emph{SAT}) attack or algebraic attack, we introduce a dynamic encryption mechanism that removes this vulnerability. We demonstrate how each communicated message in the proposed scheme reduces the level of trust. Accordingly, since a specific number of messages, $N$, could break the communication and extract the key,  by using the dynamic encryption mechanism, \emph{ExTru} can re-initiate the level of trust periodically after $T$ messages where $T<N$, to protect the communication against side-channel and scan-based attacks (e.g. SAT attack). Furthermore, we demonstrate that by properly configuring the value of $T$, \emph{ExTru} not only increases the strength of security from per “\emph{device}” to per “\emph{message}”, it also significantly improves energy consumption as well as throughput in comparison with an architecture that only uses a conventional side-channel resistant block/stream cipher.

\end{abstract}

\keywords{Internet-of-Thing, Secure Communication, Lightweight Cryptography, Block Cipher, Stream Cipher, Physical Attack}

\section{Introduction}

The Internet of Things (IoT), which has been foreseen to become the most successful business for the next decade by \emph{International Technology Roadmap for Semiconductors} (ITRS), is an inevitable landmark of smart life providing novel applications and services, ranging from business automation to personal day-to-day life \cite{gubbi2013internet,al2015internet,li2015internet}. The IoT infrastructure is the seamless connection of billions of heterogeneous devices (\emph{"things"}) within a large integrated network (the \emph{"Internet"}). The heterogeneity of IoT constitutes from a wide variety of devices, such as smartwatches, mobile phones, etc, which results in a drastic increase in the number of IoT devices, estimated to be 26 billion connected IoT devices by the end of 2020 \cite{evans2011internet}.  

Although IoT devices provide a more efficient, automated, and smart life, from security/privacy perspective, many threats and vulnerabilities have been raised in IoT devices. Many investigations on cyber-based threats demonstrate that there are 176 new cyber-threats every minute, and over 2.5 million within only four months \cite{frustaci2017evaluating}. Several incidents have highlighted the massive influence of counterfeit/cloned/tampered devices into the supply chain \cite{guin2014counterfeit,rostami2014primer}. As an instance, influencing and controlling every connected device within a ZigBee network, which is one of the most prevalent wireless communications in IoT, has been illustrated in \cite{zillner2015zigbee,ronen2017iot}. Another recent evaluation by HP demonstrates that 70\% of the devices in IoT are vulnerable to different types of threats, including physical attacks \cite{kumar2016security}.

\begin{figure}[t]
\centering
\includegraphics[width = 250pt]{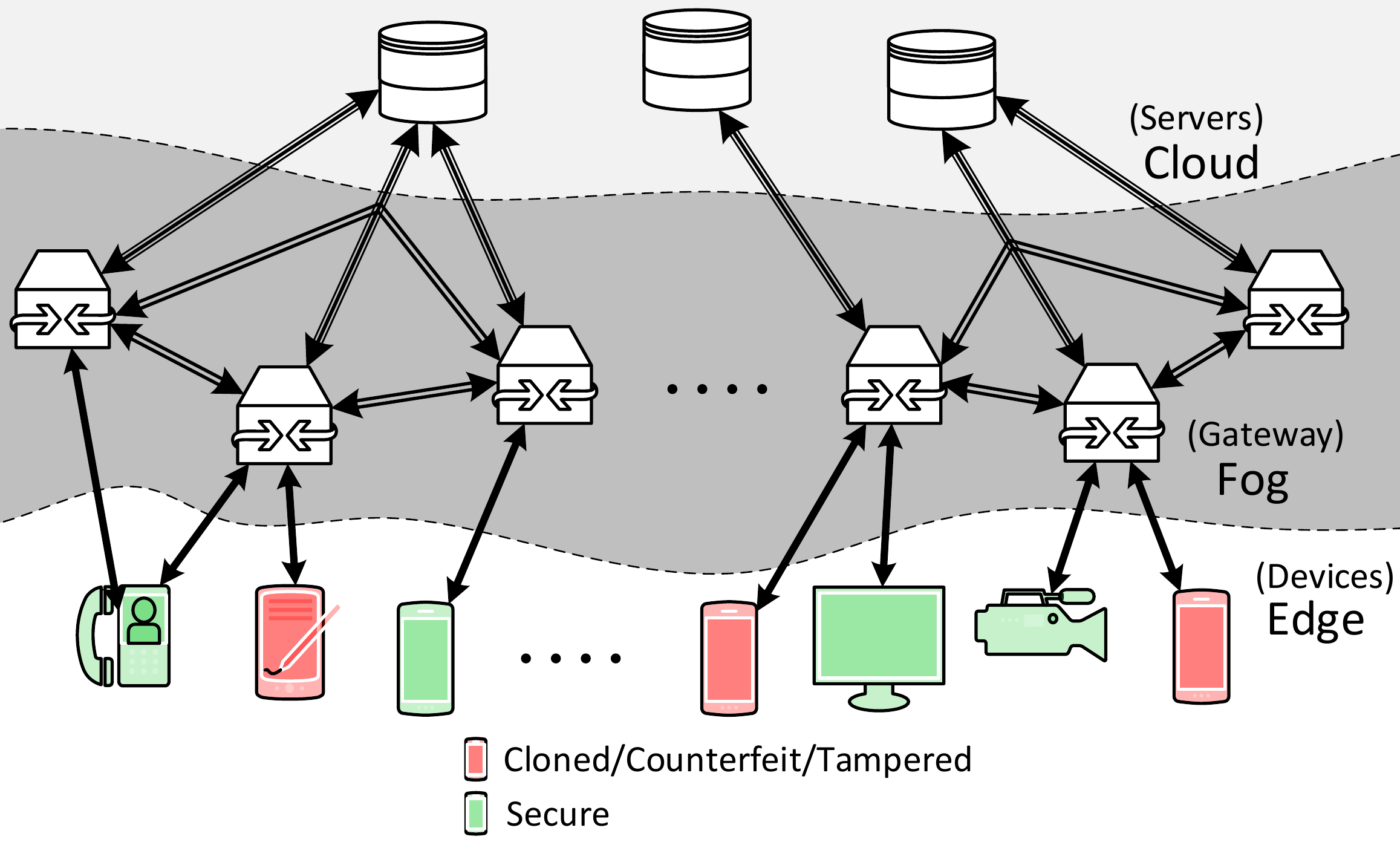}
\caption{ A Standard IoT Model with Hardware Vulnerabilities.}
\label{IoTarch}
\end{figure}

In current IoT applications, almost all proposed IoT devices are working (and communicating) based on a very well known 3-layer hierarchical architecture that is illustrated in Fig. \ref{IoTarch}. These three layers, i.e. \textbf{\emph{"devices"}}, \textbf{\emph{"gateways"}} and \textbf{\emph{"servers"}} are the main layers in IoT architecture \cite{atzori2010internet,da2014internet}. The devices that are responsible for interacting between the physical environment and computer-based systems, called \emph{edge}, can connect with servers through gateways. Accordingly, equipping edge devices with some fundamental components, including sensors, analog to digital (A/D) converters, inter-communication frameworks, memories, and embedded micro-controllers, is required, to provide the capability of collecting, processing, and relaying data in a heterogeneous network.  

Although several IoT security challenges should be considered meticulously, combating hardware threats that are generally initiated at \emph{edge} (devices layer), requires more attention \cite{frustaci2017evaluating}. Numerous solutions, including communication standards optimization, more secure configuration, etc, have been introduced to protect IoT devices and their communications against physical threats, which help to prevent the wide variety of conventional attacks \cite{pinto2017iioteed,yuan2018reliable}. For instance, the utilization of symmetric-based secret-key ciphers or keyed hash-based authentication code (HMAC) is prevalent in IoT devices to provide integrity and authentication while securely protect the inter-communication of IoT edge devices \cite{koteshwara2017comparative,shivraj2015one,kamali2016fault}.

Considering that the power consumption (particularly energy consumption) constraints in resource-constrained edge devices are very strict, the energy overhead of security solutions against hardware threats must be minimized. For instance, tight restrictions in edge devices enforce the designer to employ lightweight ciphers, such as stream ciphers or lightweight block ciphers \cite{dinu2019triathlon,beaulieu2015simon}. However, the energy consumption of this breed of encryption architectures is still high for a high portion of IoT edge devices. Also, the performance of these ciphers considerably lower than regular block ciphers. This creates an inevitable security/cost trade-off in lightweight IoT devices, which results in sacrificing one of them, i.e. the security or the cost, which motivates the research community to carry on working/investigating on a low-energy and security-enhanced communication scheme in IoT while the performance is not degraded.      

In this paper, we introduce a new lightweight, fast, and provably secure \emph{Exp}irable \emph{Tr}ust (\emph{ExTru}) mechanism relied on a configurable switching and toggling network (CSTN) as well as the winner of the Competition for Authenticated Encryption:
Security, Applicability, and Robustness (CAESAR) \cite{caesar2013competition}, called ACORN \cite{wu2016acorn}. \emph{ExTru} provably protects the inter-communication of IoT edge devices while it even obtains higher performance and mitigates the energy consumption compared to the case in which the regular block/stream ciphers have been used. Moreover, we show how \emph{ExTru} engages dynamicity in the circuit to provide guaranteed protection against different types of physical and scan-based attacks, such as side-channel, Boolean satisfiability (SAT) attack, and algebraic attack. We demonstrate that by using this dynamic encryption scheme, the strength of security could be elevated from \emph{per device} to \emph{per message}. The contributions of our paper are as follows:

\begin{enumerate}

\item By introducing a near non-blocking configurable switching and toggling network (CSTN), we show how we add dynamicity to the IoT devices intercommunication. 

\item We show that this dynamicity along with the fast and efficient ACORN invalidates the possibility of the leakage of each message, which helps to show that this approach is provably resilient against physical attacks such as side-channel, scan-based SAT attack, and algebraic attack. 

\item The dynamicity of \emph{ExTru} allows us to relax the responsibility of ACORN, which helps to considerably boost the performance of the communication channel between IoT devices while the possibility of leakage is almost \emph{ZERO}. Also by conveying part of the responsibility to the near non-blocking CSTN, we show that the energy consumption would be mitigated considerably. 

\item To depict the efficiency of \emph{ExTru} in terms of security, energy, and performance, we provide a full-detailed post-route evaluation on the proposed scheme compared to conventional IoT inter-communication mechanisms that almost use a conventional side-channel resistant block/stream.

\end{enumerate}

The rest of the paper is organized as follows: Section \ref{sec:related} presents the previous work. Section \ref{sec:proposed} elaborates the overall structure of the proposed dynamically encrypted scheme and how it is able to guarantee the security of IoT communication with significant energy mitigation as well as throughput improvement compared to conventional cipher-based communication schemes. In section \ref{attacks}, we evaluate the security of ExTru against physical attacks such as side-channel, scan-based SAT, and algebraic attack. In \ref{sec:results}, the experimental results have been provided and discussed. Finally, Section \ref{sec:conclusion} concludes the paper.

\section{Related Work}
\label{sec:related}

Due to the resource-constrained nature of IoT devices, a big challenge in guaranteeing the security of this group of devices is that the implementation of the security measures must be sufficiently lightweight, which prevents the designers to directly use conventional block ciphers, such as AES-GCM \cite{dworkin2007sp}. Many studies have been taken by the research community to not only address security issues in IoT networks but also to increase the efficiency by lowering the power (particularly energy) consumption and increasing the throughput. For instance, the fact that the elliptic curves cryptography (ECC) achieves guaranteed security with reduced resource requirements has attracted the research community \cite{piedra2013extending,nam2014provably}. The work in \cite{marin2015optimized} has constructed an optimized ECC for secure communication in heterogeneous IoT devices based on Schnorr signature. Also, a simple key negotiation protocol has been introduced in this work that is based on the Schnorr scheme to demonstrate the usability of the presented ECC optimizations. 

Based on the desirable features of a physically unclonable function (PUF), such as lightweightedness, unpredictability, unclonability, and uniqueness, many researchers have been motivated to concentrate on the usage of this module to build a secure communication for IoT devices. Among several studies on PUF-based secure communication for IoT devices \cite{halak2016overview,chatterjee2017puf,chatterjee2018building,liu2019xor}, the work in \cite{chatterjee2017puf} has introduced an authentication, key sharing, and secure communication architecture, in which each IoT device has an integrated PUF. In this work, the identity of each device is created by the challenge-response pair signature of its PUF instance, and by engaging the identity-based encryption scheme proposed in Boneh and Franklin, the security of this approach is proven against attacks like chosen-plaintext/ciphertext attack. 

Numerous software/hardware implementation of lightweight ciphers
suited for IoT devices have been proposed in recent few years, including RECTANGLE \cite{zhang2015rectangle}, PICO \cite{bansod2016pico}, Extended-LILIPUT \cite{ali2017optimised}, SIT \cite{usman2017sit}, SKINNY \cite{beierle2016skinny}, MANTIS \cite{beierle2016skinny}, to name but a few. Some of these ciphers could provide the best performance on software implementation, however, a portion of them have better performance in hardware implementation. For instance, the work in \cite{usman2017sit} introduces a lightweight 64-bit symmetric block cipher, called SIT, whose implementation is a mixture of Feistel and a uniform substitution-permutation network. The proposed approach uses some logical operations along with some swapping and substitution. Most of the encryption algorithms designed for IoT reduced the number of rounds to make a cost-security trade-off. For instance, SIT uses five rounds of encryption with 5 different keys to improve energy efficiency.

The lightweightedness of the stream ciphers, on the other hand, has received fascinated attention from many researchers’ in recent years \cite{mohd2015survey,singh2017advanced,sfar2018roadmap,manifavas2016survey}. Since IoT being an emerging field requires lightweight cipher designs with robustness, less complexity, and lower energy consumption, stream ciphers are very suited for particularly edge devices. Many studies evaluate the possibility of engaging stream ciphers in IoT devices, such as WG-8 \cite{fan2013wg}, Trivium \cite{de2005trivium}, Quavium \cite{tian2012quavium}, and ACORN \cite{wu2016acorn}. 

\section{\emph{ExTru} Infrastructure}
\label{sec:proposed}

\begin{figure}[t]
\centering
\subfloat[]{{\includegraphics[width=250pt]{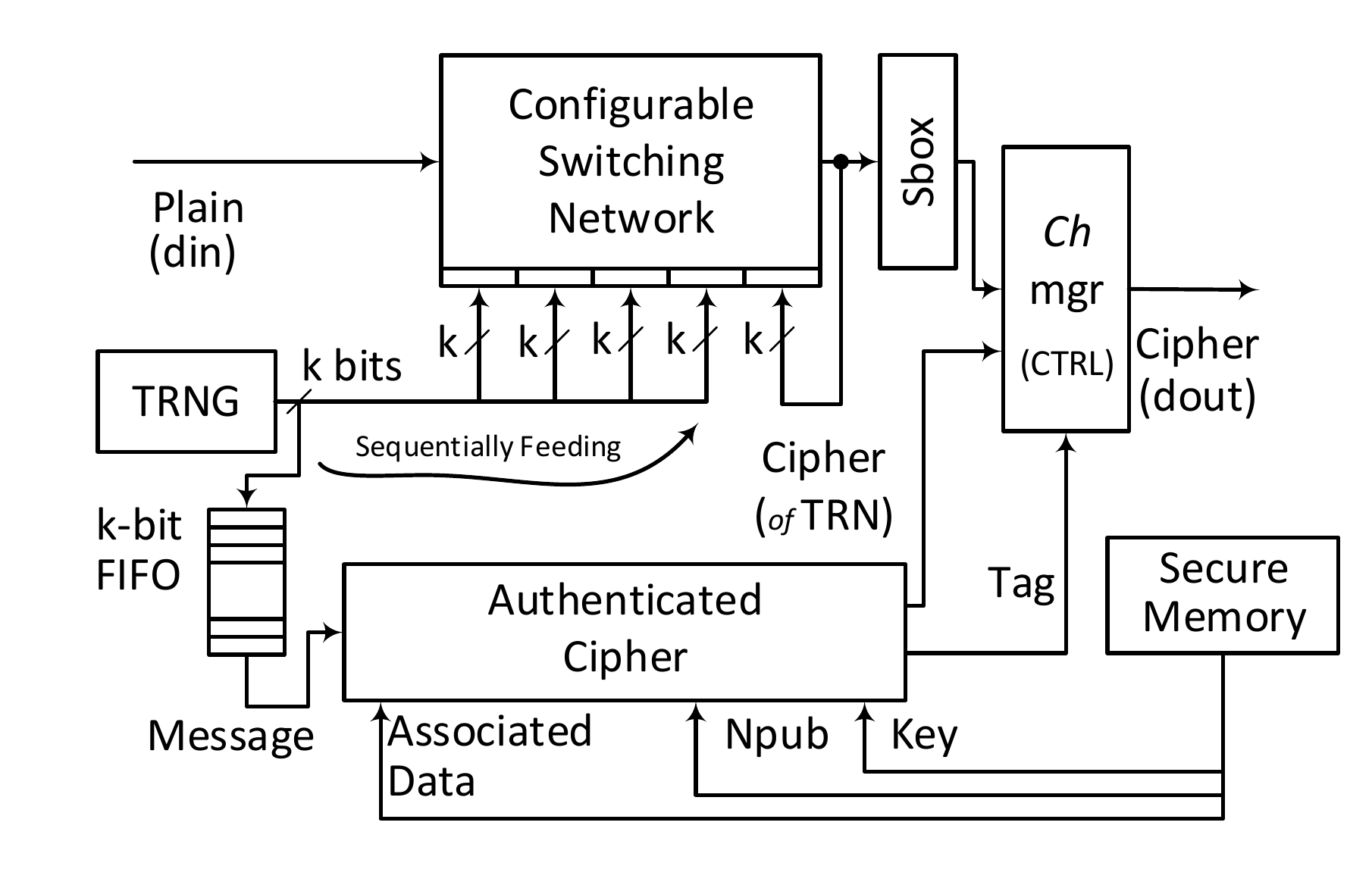} }} \newline
\subfloat[]{{\includegraphics[width=250pt]{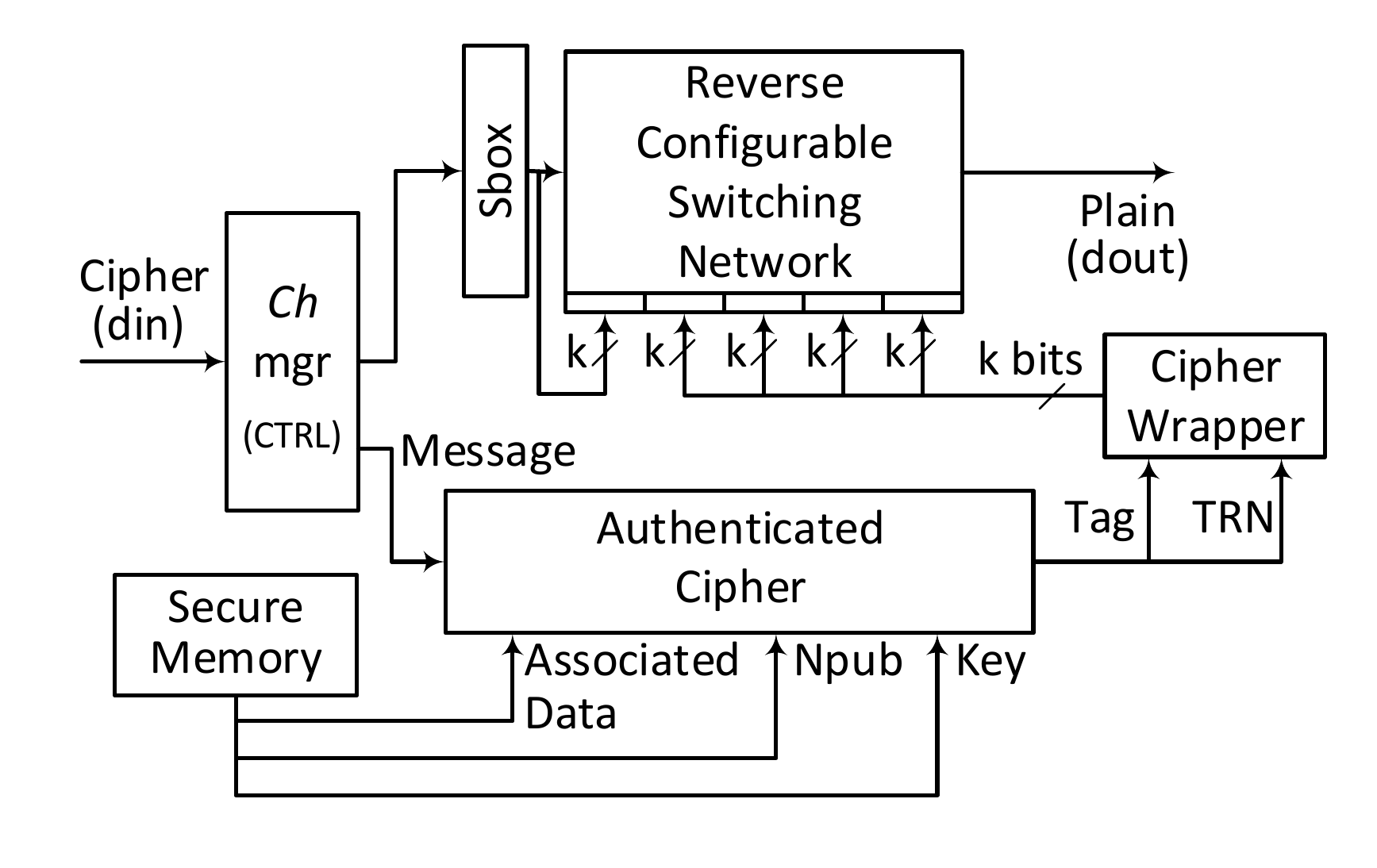} }} \newline
\caption{\emph{ExTru} Overall Infrastructure (a) Transmitter, (b) Receiver}
\label{extruarch}
\end{figure}

\emph{ExTru} consists of four main sub-modules: (1) ACORN as a stream cipher that would be used periodically (The frequency will be discussed further), (2) a configurable switching and toggling network (CSTN) that dynamically permutes/toggles the data based on the configuration generated by TRNG, (3) a random number generator (RNG) that is responsible for generating random data for Threshold Implementation of ACORN as well as for generating the CSTN configuration, and (4) a substitution box placed after CSTN to eliminate the linearity/predictability of the ciphertext.  The overall architecture of \emph{ExTru} has been demonstrated in Fig. \ref{extruarch} for both transmitter side and receiver side. 

On the transmitter side, the CSTN is used to permute/toggle the plaintext using the configuration (TRN) generated by the random number generator (RNG). The RNG will periodically change the configuration (TRN) to add dynamicity into the permutation/toggle network (CSTN). Parts of the configuration is fed by the permuted/toggled data (the output of the CSTN) to make the operation stateful (data-dependent). The CSTN is followed only by a substitution-box to eliminate the linearity/predictability of the output. The TRN that is used to configure the CSTN has been also encrypted using the authenticated cipher to be transmitted to the receiver. The key used for authenticated cipher could be pre-stored in the secure memory or produced by a PUF. The output of the transmitter (ciphertext) would be selected from the output of the s-box (permuted/toggled + substituted plaintext) or authenticated cipher output (encrypted TRN). 

On the receiver side, on the other hand, the reverse CSTN (RCSTN) must be used to recover the permuted/toggled + substituted plaintext. We will show that similar to ACORN that engages only one hardware module for both encryption/decryption, the CSTN hardware is the same for both receiving/sending operations (same hardware for both CSTN and RCSTN). Hence, no duplicated hardware (one for CSTN and one for RCSTN) is required to be added on each side. When TRN is received from the transmitter it must be decrypted using the authenticated cipher to be used as the configuration of the RCSTN. If the received data is not TRN, it first must pass the s-box to accomplish re-substitution, then it must pass the RCSTN to recover the plaintext. 

\begin{figure}[t]
\centering
\subfloat[]{{\includegraphics[width=260pt]{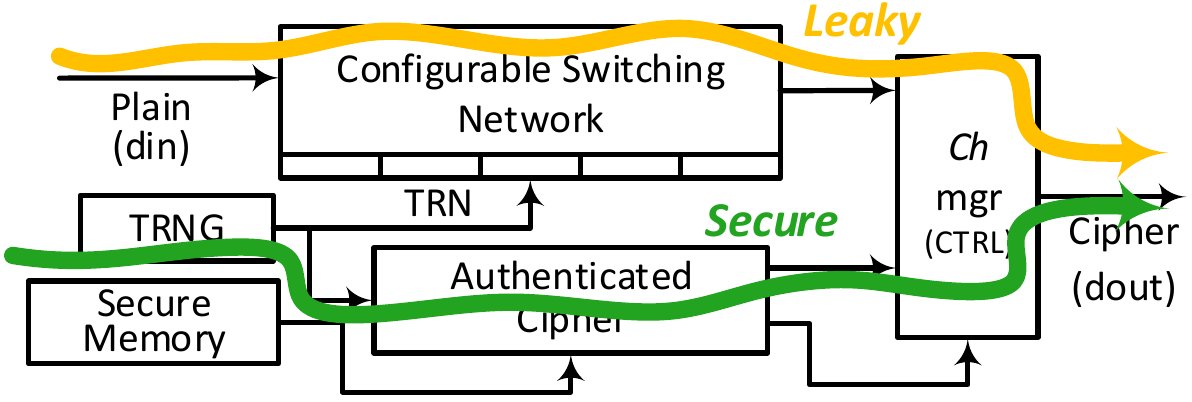} }} \newline 
\subfloat[]{{\includegraphics[width=260pt]{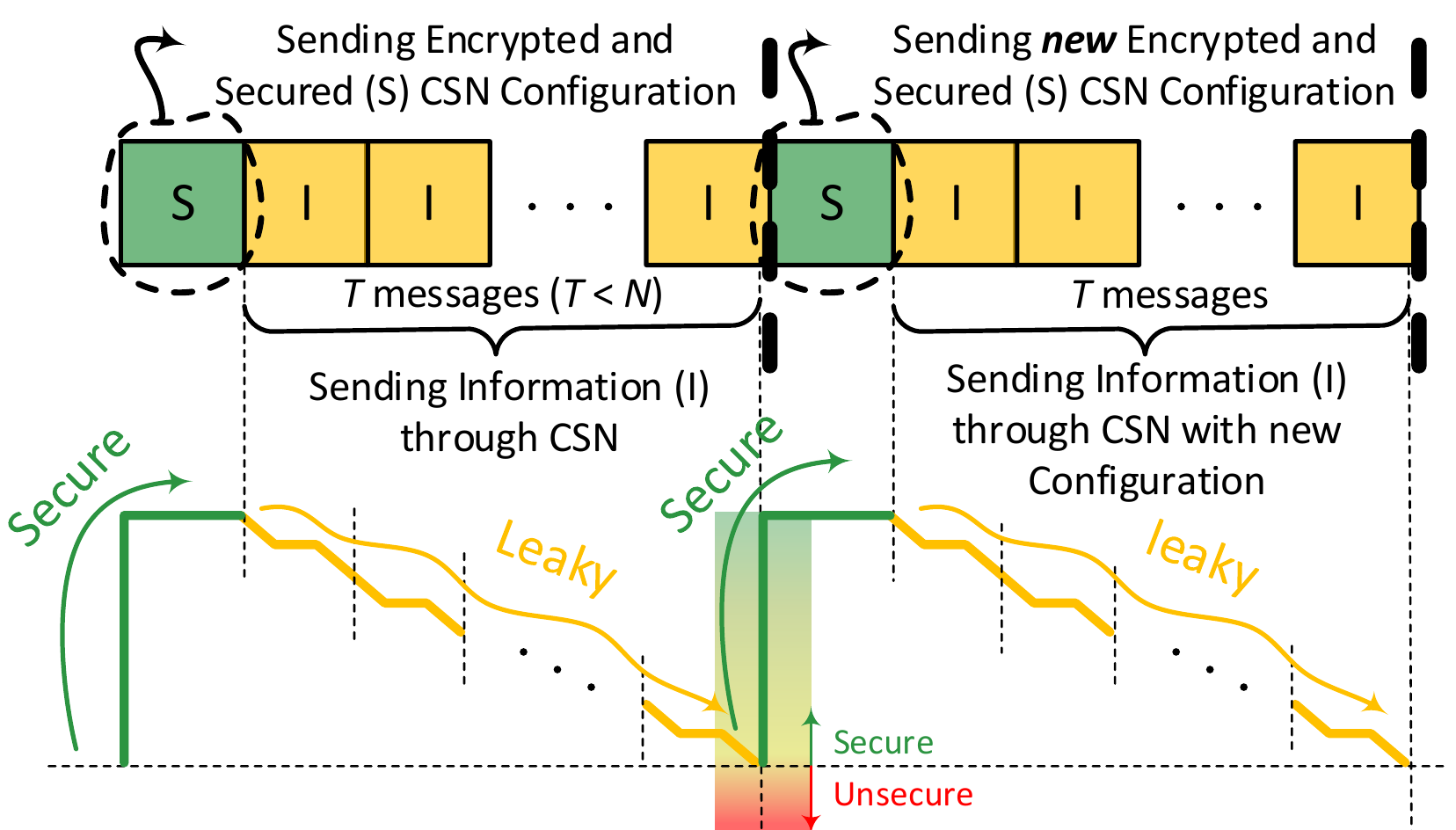} }} \newline
\caption{(a) The overall structure of using Dynamically Encrypted communication (b) Re-intensifying the Security Strength by Dynamically Encrypted communication}
\label{expdyn}
\end{figure}

Fig. \ref{expdyn} depicts the overall structure of dynamic encryption provided by \emph{ExTru}, which has no sign of leaky communication. As shown in Fig. \ref{expdyn}(b), for each specific number of transmission ($T$), which must be less than $N$, a new CSTN configuration will be sent via side-channel resistant cipher. As it is shown, a secure message ($S$), which contains TRN, will be sent periodically after every $T$ messages ($I$) that are handled by CSTN/RCSTN. Based on different forms of attacks, such as side-channel, scan-based SAT, and algebraic attack, messages ($I$) are leaky. So, periodically changing TRN ($S$) and sending through side-channel resistant ciphers re-intensify the security of the communication.

Based on the size of the CSTN/RCSTN (number of I/O), we will show that the maximum feasible update frequency ($N$) would be changed. Consequently, the CSTN configuration (TRN), which is fed by RNG, must be changed dynamically after every $T$ iterations, where $T < N$. Also, the size of CSTN/RCSTN determines the number of configuration bits (size of each $S$) must be generated by the RNG. In the following sub-sections we discuss the details of \emph{ExTru} implementation.

\subsection{Configurable Switching \& Toggling Network (CSTN)} \label{CSTN_section}

\begin{figure}[t]
\centering
\subfloat[Blocking \emph{OMEGA}]{{\includegraphics[width=280pt]{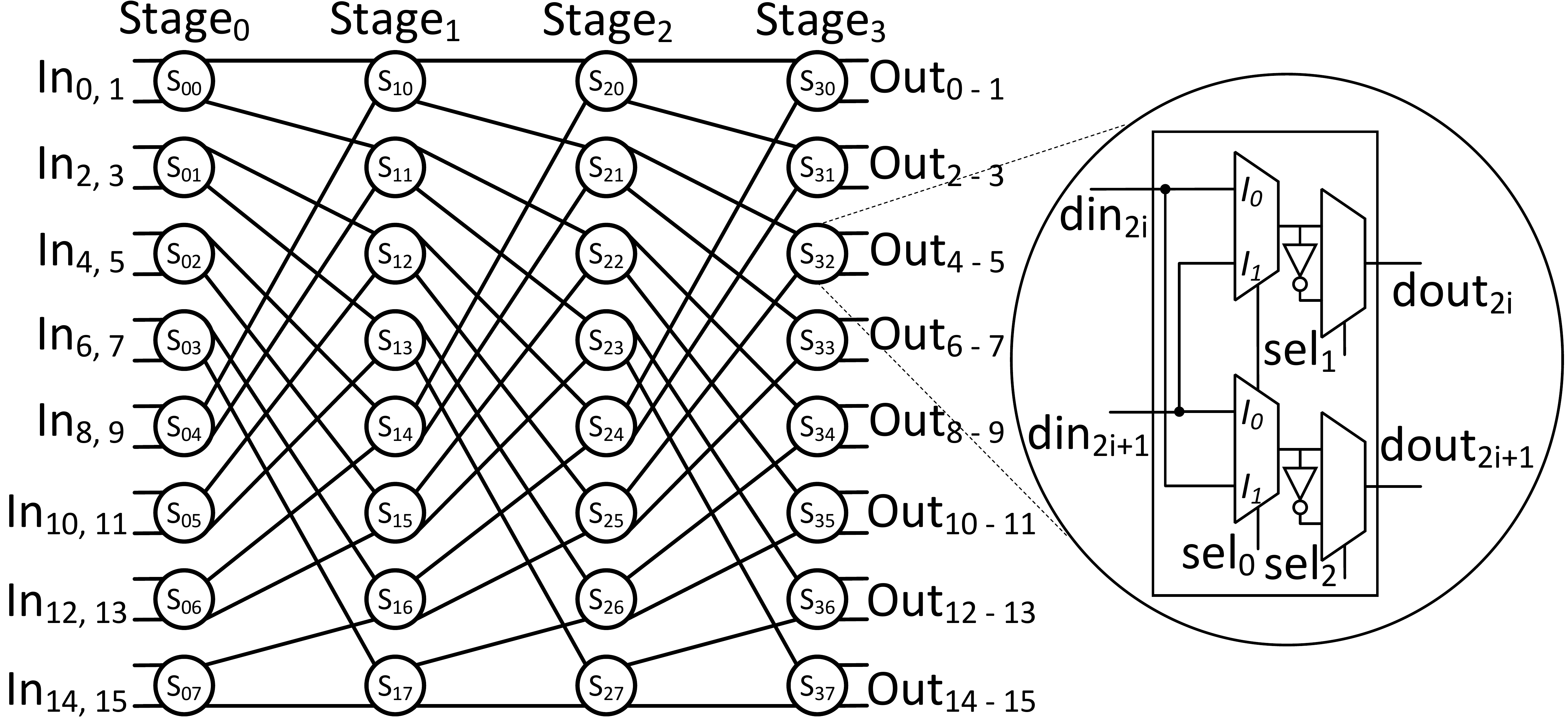} }} \newline
\subfloat[Near Non-blocking $LOG_{8, 1, 1}$]{{\includegraphics[width=280pt]{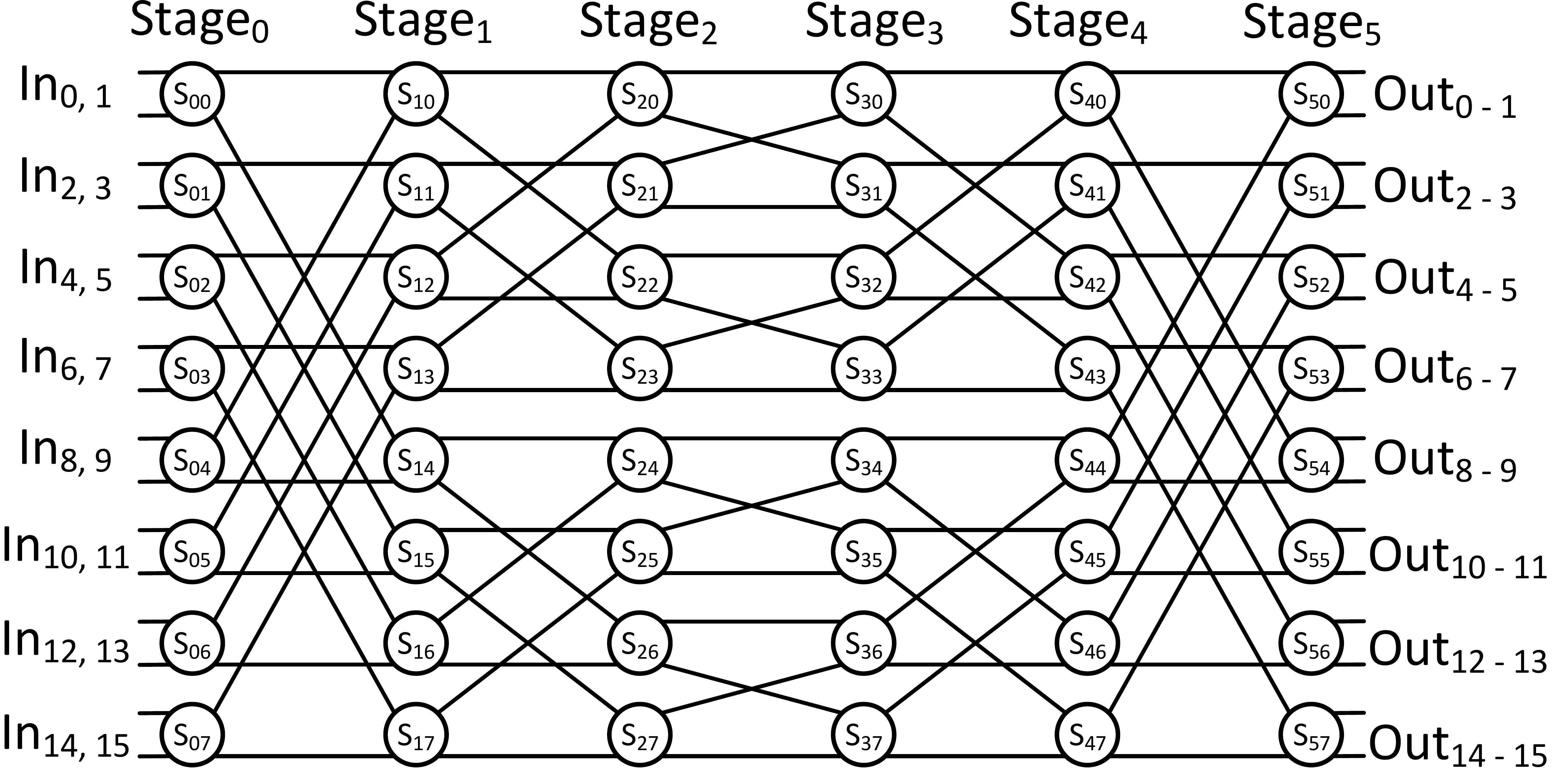} }} \newline
\caption{Logarithmic Network (a) Blocking, (b) Near Non-blocking.}
\label{CSTN_arch}
\end{figure}

The CSTN is a logarithmic routing (permutation) network that could permute the order of the signals at its input pins to its output pins while possibly toggling their logic levels based on its configuration (TRN). Fig. \ref{CSTN_arch}(a) captures a simple implementation of an 8$\times$8 CSTN based on \emph{OMEGA} network \cite{ahmadi1989survey}. The network is constructed using permutation elements, denoted as Re-Routing Blocks (RRB). Each RRB is able to possibly toggle and permute each of the input signals to each of its outputs. The number of RRBs needed to implement this simple CSTN for $N$ inputs ($N$ is a power of 2) is simply $N/2\times logN$.

Each CSTN should be paired with an RCSTN. RCSTN must be able to reverse all operations accomplished by CSTN to re-generate the plaintext. Due to the structure of CSTN, RCSTN can be implemented by \emph{vertically flipping} the CSTN without any change in configuration \cite{goke1973banyan}. In fact, by vertically flipping the CSTN, and then applying the same configuration, we re-generate plaintext. So, implementing RCSTN by vertically flipping the CSTN allows us to use the same configuration for both CSTN and RCSTN. However, to avoid duplicating the hardware (to put one dedicated hardware for CSTN and one dedicated hardware for RCSTN), by flipping the configuration bits (row-pivot reversed TRN), the CSTN would operate as its corresponded reverse CSTN. Hence, only one hardware is enough to operate as both CSTN and RCSTN (using TRN or row-pivot reversed TRN).

The \emph{OMEGA} network along with many other networks of such nature (\emph{BUTTERFLY}, etc.) are blocking networks \cite{ahmadi1989survey}, in which we cannot produce all permutations of input at the network's output pins. This limitation significantly reduces the ability of a CSTN to randomize its input. Also, Evaluation of this permutation networks as a means of obfuscation to defend supply chain shows that the blocking version of this breed of networks could be easily broken by a SAT attack within few iterations \cite{kamali2019full,azar2019coma}. 

Being a blocking or a non-blocking CSTN depends on the number of stages in CSTN. Since no two paths in an RRB are allowed to use the same link to form a connection, for a specific number of RRB columns, only a limited number of permutations is feasible. However, adding extra stages could transform a blocking CSTN into a strictly non-blocking CSTN. Using a strictly non-blocking CSTN not only improves the randomization of propagated messages through the CSTN, but also improves the resiliency of these networks against possible SAT attacks for extraction of a TRN used as the key for a CSTN-RCSTN cipher. A non-blocking logarithmic network could be represented using $LOG_{n, m, p}$, where $n$ is the number of inlets/outlets, $m$ is the number of extra stages, and $p$ indicates the number of copies \emph{vertically cascaded} \cite{shyy1991log}. 

According to \cite{shyy1991log}, to have a strictly non-blocking CSTN for an arbitrary $n$, the smallest feasible values of $p$ and $m$ impose very large area/power overhead. For instance, for $n=64$, the smallest feasible values, which make it strictly non-blocking, are $m=3$ and $p=6$, which means there exists more than $5\times$ as much overhead compared to a blocking CSTN with the same $n$, resulting in a significant increase in the area and delay overhead. To avoid such large overhead, we employ a \emph{close to non-blocking CSTN} described in \cite{shyy1991log} to implement the CSTN-RCSTN pair. This network is able to generate not all, but \emph{almost all} permutations, while it could be implemented using a $LOG_{n, log_2(n) - 2, 1}$ configuration, meaning it needs $log_2(n) - 2$ extra stages and no additional copy. Fig. \ref{CSTN_arch}(b), demonstrates an example of such a near non-blocking CSTN with $n = 8$.

Based on the structure of CSTN/RCSTN, and the size used for implementation, the size of configuration bits ($S$) would be changed. For instance, for a near non-blocking $LOG_{64, 4, 1}$, the number of selectors is 960 $(2log_2(64) - 2)(32)(3)$ (3 selectors in each $2 \times 2$ switches (Fig. \ref{CSTN_arch})). Based on the size of configuration, and the number of messages that could be sent in each interval ($T$), the overhead (time/energy) would be changed in \emph{ExTru}. However, we show that since ($T$) is large enough, the performance boost, as well as the mitigating of the energy consumption, would be considerably high. 

\subsection{Authenticated Encryption with Associated Data} \label{AEAD}

The Authenticated Encryption with Associated Data (AEAD) is used in \emph{ExTru} for the transmission of the CSTN-RCSTN configuration (TRN). Authenticated ciphers incorporate the functionality of confidentiality, integrity, and authentication. The input of an authenticated cipher includes \emph{plaintext} (message), \emph{associated data} (AD), \emph{public message number} (NPUB), and \emph{secret key}. Then, the \emph{ciphertext} is generated as a function of these inputs. A \emph{tag}, which depends on all inputs, is generated after message encryption to assure the integrity and authenticity of the transaction. This tag is then verified after the decryption process. The choice of AEAD could significantly affect the area overhead of the solution, the speed of encrypted communication, and the extra energy/power consumption. To show the performance, power/energy, and area trade-offs, we employ two AEAD solutions: a NIST compliant solution (AES-GCM) \cite{dworkin2007sp}, and a promising lightweight solution (ACORN) \cite{wu2016acorn}.

AES-GCM is the current National Institute of Standards and Technology (NIST) standard for authenticated encryption and decryption as defined in \cite{dworkin2007sp}. ACORN is one of two finalists of the Competition for Authenticated Encryption: Security, Applicability, and Robustness (CAESAR), in the category of lightweight authenticated ciphers, as defined in \cite{wu2016acorn}. An 8-bit side-channel protected version of AES-GCM and a 1-bit side-channel protected version of ACORN are implemented as described in \cite{diehl2018face}. Both implementations comply with lightweight version of the CAESAR HW API \cite{homsirikamol2015gmu}. 

Our methodology for side-channel resistant is threshold implementation (TI), which has wide acceptance as a provably secure Differential Power Analysis (DPA) countermeasure~\cite{nikova2006threshold}. In TI, sensitive data is separated into shares and the computations are performed on these shares independently. TI must satisfy three properties: (1) Non-completeness: Each share must lack at least one piece of sensitive data, (2) Correctness: The final recombination of the result must be correct, and (3) Uniformity: An output distribution should match the input distribution. To ensure uniformity, we refresh TI shares after non-linear transformations using randomness. We use a hybrid 2-share/3-share approach, where all linear transformations in each cipher are protected using two shares, which are expanded to three shares only for non-linear transformations. 

To verify the resistance against DPA, we employ the Test Vector Leakage Assessment methodology in \cite{gilbert2011testing}. We leverage a "fixed versus random" non-specific t-test, in which we randomly interleave first fixed test vectors and then randomly-generated test vectors, leading to two sequences with the same length but different values. Using means and variances of power consumption for our fixed and random sequences, we compute a figure of merit $t$. If $|t| > 4.5$, we reason that we can distinguish between the two populations and that our design is leaking information. The protected AES-GCM design has a 5-stage pipeline and encrypts one 128-bit input block in 205 cycles. This requires 40 bits of randomness per cycle. In ACORN-1, there are ten 1-bit TI-protected AND-gate modules, which consume a total of 20 random re-share, and 10 random refresh bits per state update. In a two-cycle architecture, 15 random bits are required per clock cycle.

\subsection{\textbf{Random Number Generator (RNG)}} \label{RNG}

A RNG unit is required on both sides to generate random bits for side-channel protection of AEAD units, a random public message number (NPUB) for AEAD, and TRNs for CSTN-RCSTN. We adopted the ERO TRNG core described in \cite{petura2016survey}, which is capable of generating only 1-bit of random data per over 20,000 clock cycles. In our TI implementations, AES-GCM needs 40 and ACORN 15 bits of random data per cycle. So, we employed a hybrid RNG unit combining the ERO TRNG with a Pseudo Random Number Generator (PRNG). TRNG output is used as a 128-bit seed to PRNG. The PRNG generates random numbers needed by other components. The reseeding is performed only once per activation.

We adopted two different implementations of PRNG: (1) AES-CTR PRNG, which is based on AES, is compliant with the NIST standard SP 800-90A, and generates 12.8 bits per cycle. (2) Trivium based PRNG, which is based on the Trivium stream cipher described in \cite{de2005trivium}. The Trivium-based PRNG is significantly smaller in terms of area and much faster than AES-CTR PRNG. It can generate 64 bits of random data per cycle, however, it is not compliant with the NIST standard.

Also, the ERO TRNG is equipped with standard-statistical-tests applied post-fabrication, such as Repetition-Count test and the Adaptive-Proportion test, as described in NIST SP 800-90B \cite{barker2012recommendation}, any attempt at weakening the TRNG during regular operation (i.e. fault attack) can be detected by continuously checking the output of a source of entropy for any signs of a significant decrease in entropy, noise source failure, and hardware failure.

\subsection{\textbf{Substitution Box (S-Box)}} \label{sbox}

To eliminate the linearity/predictability in \emph{ExTru}, a non-feistel
trial strategy has been used that is based on Khazad block cipher \cite{barreto2000khazad}. The wide trial strategy is composed of several linear and non-linear transformations that ensures the dependency of output bits on input bits  in a complex manner \cite{daemen1995cipher}. The input and output correlation of this strategy is very large if the linear approximation is done for even one round. Also the transformation is kept uniform which treats every bit in a similar manner and provides opposition to differential attacks. 

\section{Security Analysis of \emph{ExTru}}  \label{attacks}

Assuming that the attacker can monitor the side-channel information of the chips during normal operations (based on power/current traces), and the possibility of having access to the scan chain to apply any form of scan-based attack, in this section we evaluate the resiliency of \emph{ExTru} against different physical attacks, such as side-channel, the scan-based SAT, and algebraic attack. An Attack objective may be (1)  extracting the secret key, or (2) extracting CSTN configuration (TRNs), or (3) eavesdropping on messages exchanged between the devices. 

\begin{figure}[t]
\centering
\subfloat[AES-GCM unprotected]{{\includegraphics[width=\textwidth/2]{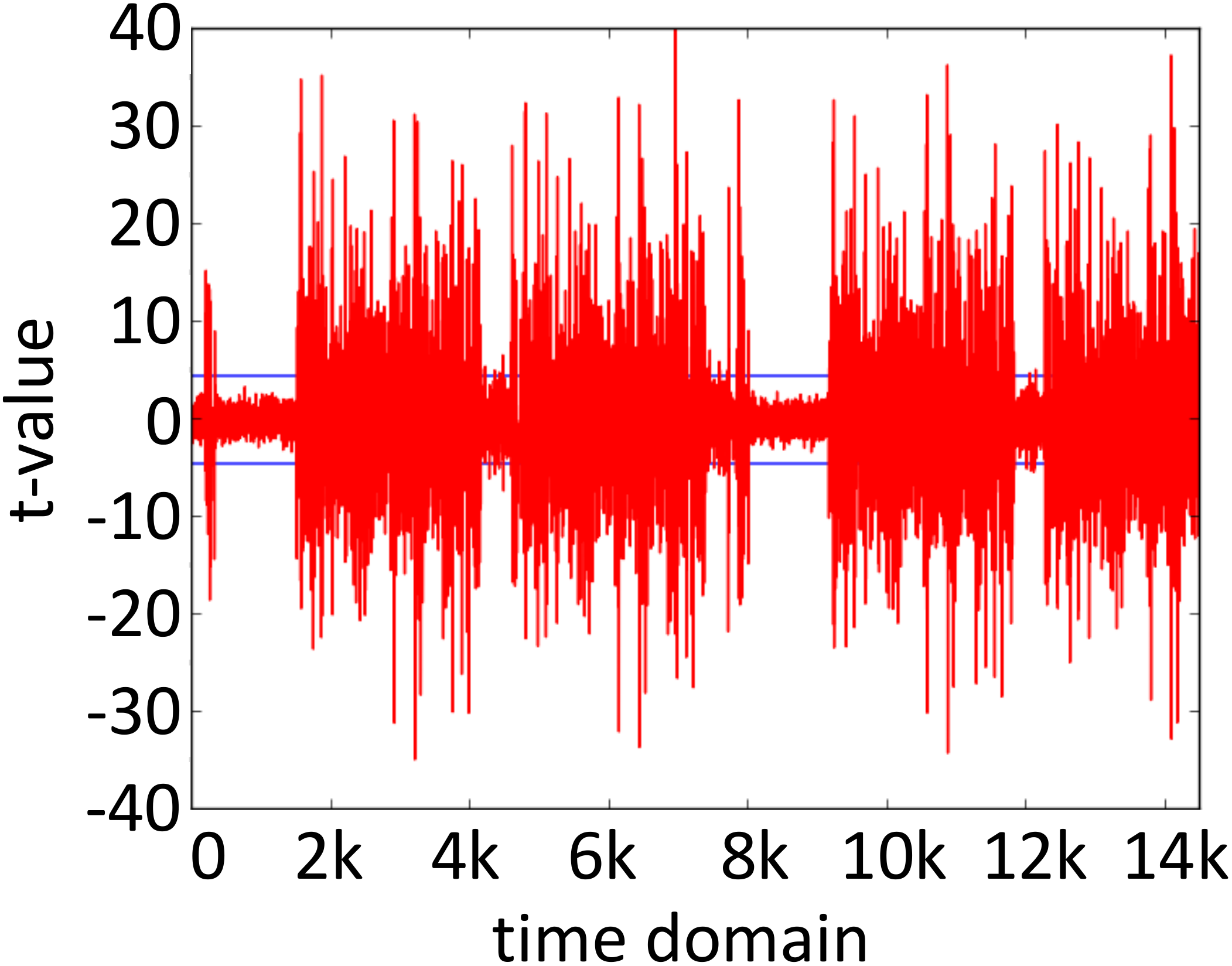} }} 
\subfloat[AES-GCM protected]{{\includegraphics[width=\textwidth/2]{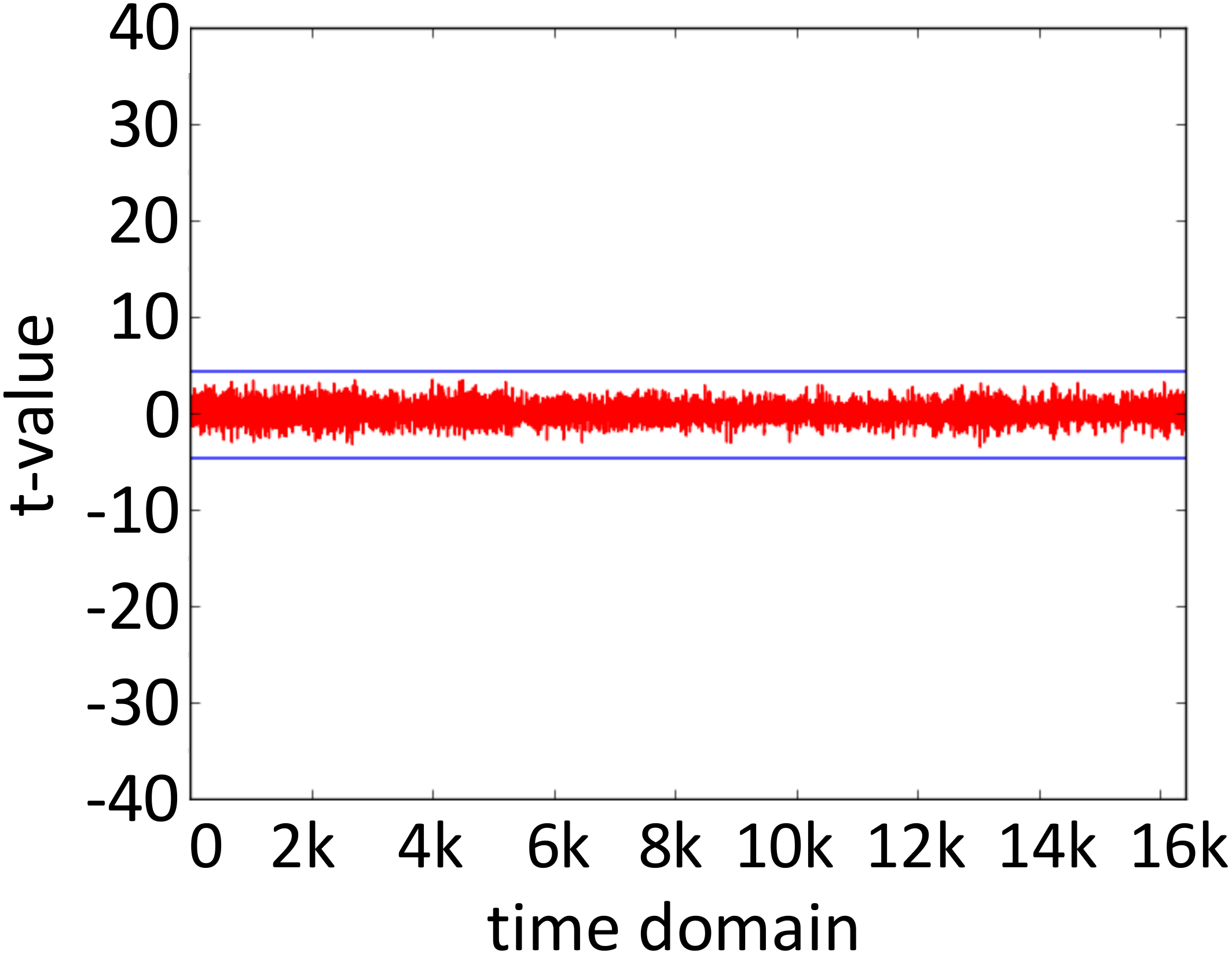} }} \\
\subfloat[ACORN unprotected]{{\includegraphics[width=\textwidth/2]{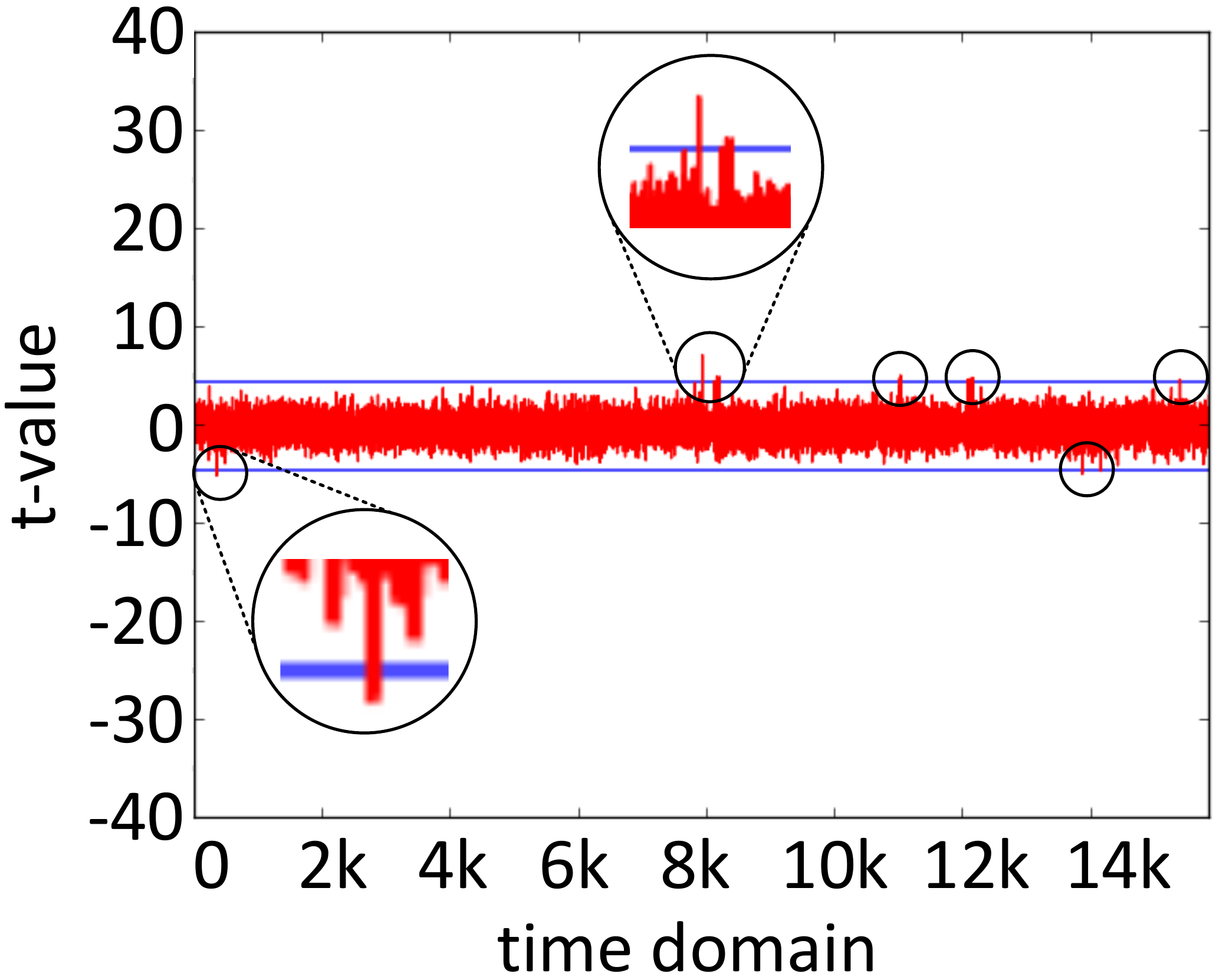} }} 
\subfloat[ACORN protected]{{\includegraphics[width=\textwidth/2]{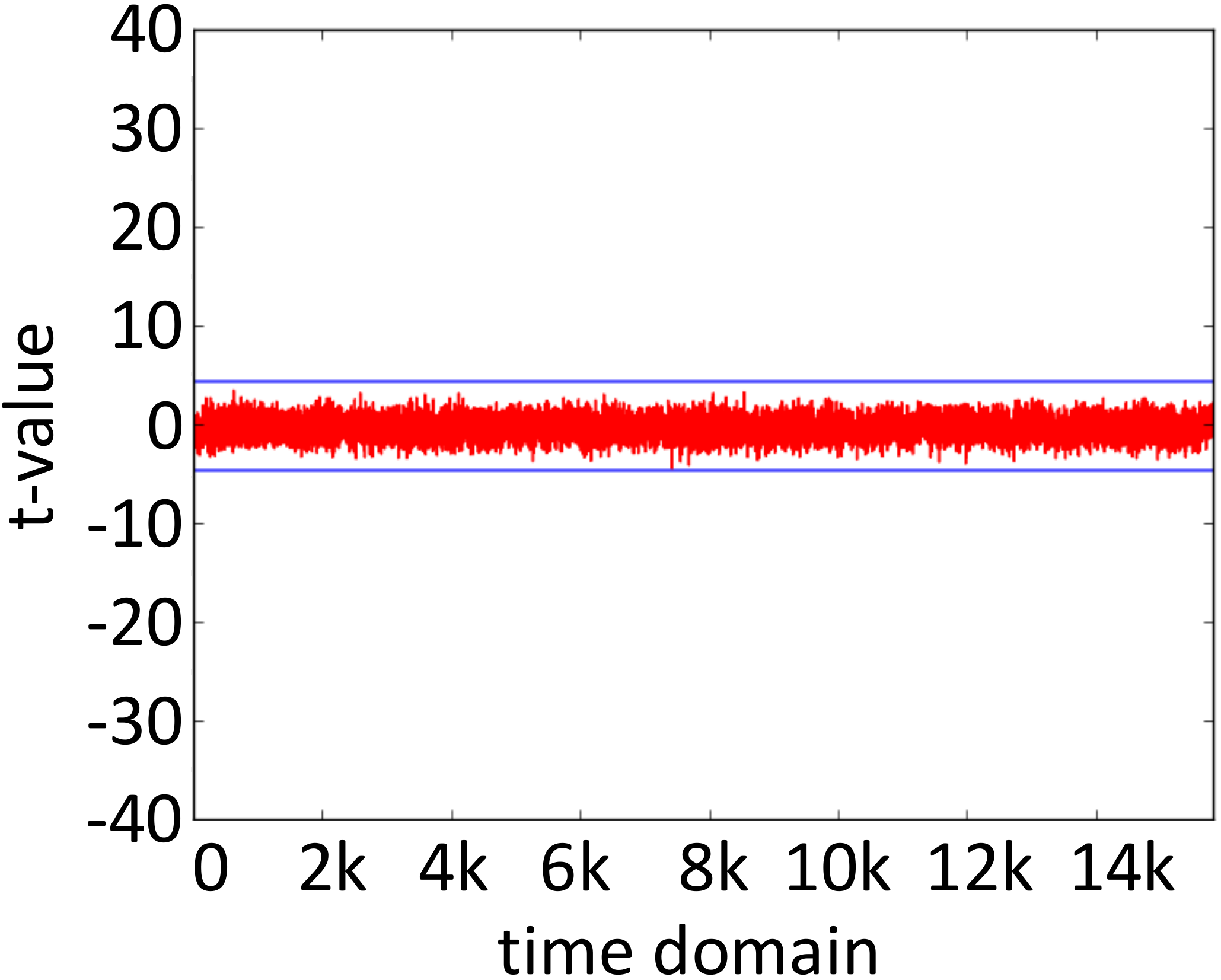} }} 
\caption{The t-test results for unprotected and protected implementations of AES-GCM and ACORN.}
\label{ttest}
\end{figure}

\subsection{\textbf{Side-Channel Attack (SCA)}}

The objective of SCA on \emph{ExTru} is to extract either the secret key used by AEAD (ACORN) or the TRN used by CSTN. Extracting a secret key is sufficient to break the communication. By extracting the secret key, the attacker can decrypt the TRN transmitted between transmitter/receiver, and by knowing the TRN, the plaintext could be recovered. Similarly, extracting the TRN reveals the communicated messages, however, since the TRN would be updated dynamically, extracting the TRN would reveal only part of the messages. It is worth mentioning that assuming that the secret key or TRN is extracted, the functionality of the s-box would be revealed using specific messages. 

Fig. \ref{ttest} captures our assessment of the side-channel resistance of AEAD using a t-test for unprotected and protected implementations of AES-GCM and ACORN  \cite{diehl2018comparison}. As illustrated, both implementations pass the t-test, indicating the guaranteed resistance against SCA. Note that this guaranteed resistance against SCA shows the robustness of communication channel during TRN transmission. 

In addition, by adding the dynamicity in \emph{ExTru}, any form of attacks, including SCA, the SAT, and algebraic, must be carried out in a limited time while the TRN of the CSTN/RCSTN is unchanged. As soon as the TRN is renewed, the previous side-channel traces or SAT iterations or algebraic calculations are useless. The period of TRN updates introduces a trade-off between energy and security and can be pushed to maximum security by changing the TRN for every new input. 

\subsection{\textbf{TRN Extraction using the SAT attack}} 

Since the attacker might have access to the scan chain to apply any form of scan-based attack, it might be possible to recover and extract the TRN by applying specific inputs to the CSTN and observing the output. This could be done by using the SAT attack that is a very applicable and known attack on logic locking schemes \cite{zamiri2019threats}\nocite{azar2019smt}. In this scheme, assuming that the TRN is the unknown parameters (such as key in logic locking), based on Table \ref{omega_sat}, it is evident that using blocking CSTN, particularly small size CSTN, does not make the design resilient against the SAT attacks. The number of iterations in Table \ref{omega_sat} shows the number ($N$) of specific inputs identified by SAT solver, called Discriminating Inputs (DIPs) \cite{subramanyan2015evaluating}\nocite{roshanisefat2018srclock}\nocite{kamali2018lut}. Finding $N$ DIPs by SAT solver allows the attacker to find CSTN/RCSTN configuration (TRN), and consequently breaks the scheme. It is evident that increasing the size of CSTN will increase $N$ (e.g. from $N=6$ in size 4 to $N=25$ in size 256). For an \emph{OMEGA}-based CSTN with size 512, SAT is not able to find the TRN after $2\times10^6$ seconds. Even after $2\times10^6$ seconds execution of SAT, it could find only 7 DIPs. However, we expect that for an \emph{OMEGA}-based CSTN with size 512, SAT needs more than 25 DIPs to find TRN.

\begin{table}[t]
\footnotesize
\centering
\caption{SAT Execution Time on \emph{OMEGA}-based Blocking CSTN for Different Sizes.}
\label{omega_sat}
\setlength\tabcolsep{1.5pt} 
\begin{tabular}{@{} l *9c @{}}
\toprule
CSTN Size ($n$)                       &   4   &   8   &   16  &   32  &   64    &   128  &   256   &   512 \\ 
\midrule
SAT Iterations                  &   6   &   7   &   8   &   12  &   14    &   24   &   25    &   TO \\
\midrule
SAT Execution Time $_{(Seconds)}$    &   0.01& 0.03  & 0.2   &  0.8  &   5.9   &  130.5 &  1136.2 &   TO \\ \bottomrule
\textit{TO: Timeout = $2\times10^6$ seconds}
\end{tabular}
\end{table}

Table \ref{nonblk_sat} illustrates that using near non-blocking CSTN considerably enhances the resiliency of this approach against the SAT attack. As shown in Table \ref{nonblk_sat}, for a near non-blocking CSTN with a size of 64 ($LOG_{64, 4, 1}$), the SAT is not able to find the TRN after $2\times10^6$ seconds. Even after $2\times10^6$ seconds execution of SAT, it cannot find more than 5 DIPs. However, based on the SAT iterations for $LOG_{32, 3, 1}$, we expect that for a close to non-blocking CSTN with size 64, more than 32 DIPs are required to extract CSTN configuration. 

\begin{table}[t]
\footnotesize
\centering
\caption{SAT Execution Time on a Close to Non-blocking CSTN, $LOG_{n, log_2(n) - 2, 1}$, for Different Sizes.}
\label{nonblk_sat}
\setlength\tabcolsep{5pt} 
\begin{tabular}{@{} l *9c @{}}
\toprule
CSTN Size ($n$)                        &   4   &   8   &   16  &   32  &   64   \\ 
\midrule
SAT Iterations                  &   14   &   18   &   25   &   32  &   TO  \\
\midrule
SAT Execution Time $_{(Seconds)}$    &   0.01& 0.015  & 2.35   &  79.18  &  TO \\ \bottomrule
\textit{TO: Timeout = $2\times10^6$ seconds}
\end{tabular}
\end{table}

\subsection{\textbf{Algebraic Attacks}} 

Algebraic attacks involve (a) expressing the cipher operations as a system of equations, (b) substituting in known data for some variables, and (c) solving for the key. ACORN has been demonstrated to be resistant against all known types of algebraic attacks, including linear cryptanalysis. Therefore, in the absence of any new attacks, the TRN transmission mode is resistant against algebraic attacks. Using CSTN and RCSTN by itself is new and requires more analysis. CSTN can be expressed as an affine function of the data input $x$, of the form $y=A\cdot x + b$, where $A$ is an $n \times n$ matrix and $b$ is an $n \times 1$ vector, with all elements dependent on the input TRN. Although recovering $A$ and $b$ is not equivalent to finding the TRN, it may enable the successful decryption of all blocks encrypted using a given TRN. We protect against this threat in numerous ways: (1) The number of blocks encrypted using a given TRN is set to the value smaller than $n$, which prevents generating and solving a system of linear equations with $A$ and $b$ treated as unknowns, (2) a part of the configuration is data-dependent and is fed from the output of the CSTN (stateful), so the values of $A$ and $b$ are not the same in any two encryptions, without the need of feeding CSTN with two completely different TRN values, (3) the substitution box added after the CSTN will eliminate all linearity/predictability of the CSTN using the algebraic attack.  

\section{Experimental Setup and Analysis}
\label{sec:results}

For evaluation, all designs have been implemented using Verilog HDL, and have been synthesized for both FPGA and ASIC targets. For ASIC verification, we used Synopsys generic 32nm process. For FPGA verification, we targeted a small FPGA board, Digilent Nexys-4 DDR with Xilinx Artix 7 (XC7A100T-1CSG324). In addition, for SAT evaluation, we employed the Lingling-based SAT attack \cite{subramanyan2015evaluating} on a Dell PowerEdge R620 equipped with Intel Xeon E5-2670 2.6 GHz and 64GB of RAM. Also, as noted, a run-time limit of $2 \times 10^6$ seconds was set for the SAT solver. For ciphers, we used two side-channel resistant ciphers (AES-GCM128 as a block authenticated cipher, and ACORN as a lightweight stream cipher). We have two modes in \emph{ExTru}: (1) \emph{ExTru} with AES-GCM, compared with its corresponding cipher (AES-GCM), (2) \emph{ExTru} with ACORN, compared with its corresponding cipher (ACORN). All configurations are listed in Table \ref{extru_config}.

\begin{table}[t]
\footnotesize
\centering
\caption{Main features of the two ExTru Modes.}
\label{extru_config}
\setlength\tabcolsep{5pt} 
\begin{tabular}{@{} l  *3c @{}}
\toprule
\multicolumn{1}{c}{Feature} & Block & Stream \\
\midrule
AEAD    & AES-GCM & ACORN \\
PRNG    & AES-CTR & Trivium \\
BUS Width & 8 & 8 \\
Pins used for Communication & 8 & 8 \\
CSTN-RCSTN Size & 64 & 64 \\
Trusted Memory & 4 Kbits & 4 Kbits \\
C$_{fix}$: initialization overhead (cycles) & 10,492 & 20,452 \\
C$_{byte}$: cycles needed for encrypting each byte & 72 & 17  \\ 
PRNG$_{perf}$: Throughput of generating TRN & $128bit / 10cycles$ & $64bit / cycle$  \\ \bottomrule
\end{tabular}
\end{table}

Table \ref{basic_ppa} demonstrates the resource utilization of $LOG_{64, 4, 1}$ compared to both ciphers using Synopsys generic 32nm library, after post-layout (route) verification (PLS). As it can be seen, PLS reports show that the power consumption of $LOG_{64, 4, 1}$ is higher than ACORN. However, based on the area utilization, $LOG_{64, 4, 1}$ is considerably smaller than ACORN and AES-GCM. The main reason is that the switching activity of CSTN is high due to numerous permutation/toggling + substitution which leads to have higher power consumption than ACORN. Additionally, the delay of critical paths in both ciphers is higher than that of CSTN. Based on Fig. \ref{extruarch}, it is obvious that critical path in \emph{ExTru} is same as that of its corresponding cipher. Consequently, we expect that the delay of critical path in \emph{ExTru} is approximately equal with that of ciphers. 

Also, Table \ref{blk_vs_nonblk} depicts area, power, and the delay of CSTNs in both blocking and near non-blocking mode with different sizes in the Synopsys generic 32nm process. As shown, it is evident that using a close to non-blocking CSTN with size 64, $LOG_{64, 4, 1}$, provides the most efficient CSTN structure, which is resilient against SAT attack. It should be noted that due to having extra stages in close to non-blocking CSTNs, the delay of these networks is slightly higher than the blocking CSTNs with the same $n$, which is negligible.

\begin{table}[t]
\footnotesize
\centering
\caption{Power, Area, and Delay of $LOG_{64, 4, 1}$ compared to protected AES-GCM and ACORN \cite{diehl2018face}}
\label{basic_ppa}
\setlength\tabcolsep{5pt} 
\begin{tabular}{@{} l *9c @{}}
\toprule
\multicolumn{1}{c}{Design} & & & Power ($uW$)  & & & Area ($nm^2$)& & & Delay ($ns$) \\
\midrule
$LOG_{64, 4, 1}$    & & & 1625.5   & & & 9965.9  & & & 1.74 \\ 
AES-GCM             & & & 3587.1    & & & 102487.5   & & & 2.48 \\
ACORN              & & & 880.9   & & & 21843.4  & & & 2.3 \\ \bottomrule
\end{tabular}
\end{table}

\begin{table}[t]
\footnotesize
\centering
\caption{SAT Execution Time on Close to Non-blocking CSTN, $LOG_{n, log_2(n) - 2, 1}$, for Different Sizes.}
\label{blk_vs_nonblk}
\setlength\tabcolsep{5pt} 
\begin{tabular}{@{} l *9c @{}}
\toprule
\multicolumn{1}{c}{CSTN} &   Area ($nm^2$)  &   Power ($uW$)   &   Delay ($ns$)  &   SAT-Resilient    \\
\midrule
omega32        &   1013.1    &   44.8   &   1.12 & \xmark  \\
log(32, 3, 1)  &   3067.5    &   213.5  &   1.33 & \xmark  \\
omega64        &   2285.5    &   107.1  &   1.22 & \xmark  \\
\textbf{log(64, 4, 1)}  &   \textbf{7438.8}    &   \textbf{845.1}  &   \textbf{1.73} & \textbf{\checkmark}  \\
omega128       &   5081.5    &   250.3  &   1.25 & \xmark  \\
omega256       &   11364.9   &   579.1  &   1.35 & \xmark  \\
\textbf{omega512}       &   \textbf{25458.3}   &   \textbf{2308}   &   \textbf{1.42} & \textbf{\checkmark}  \\
\bottomrule
\end{tabular}
\end{table}

\begin{figure}[t]
\centering
\includegraphics[width = 310pt]{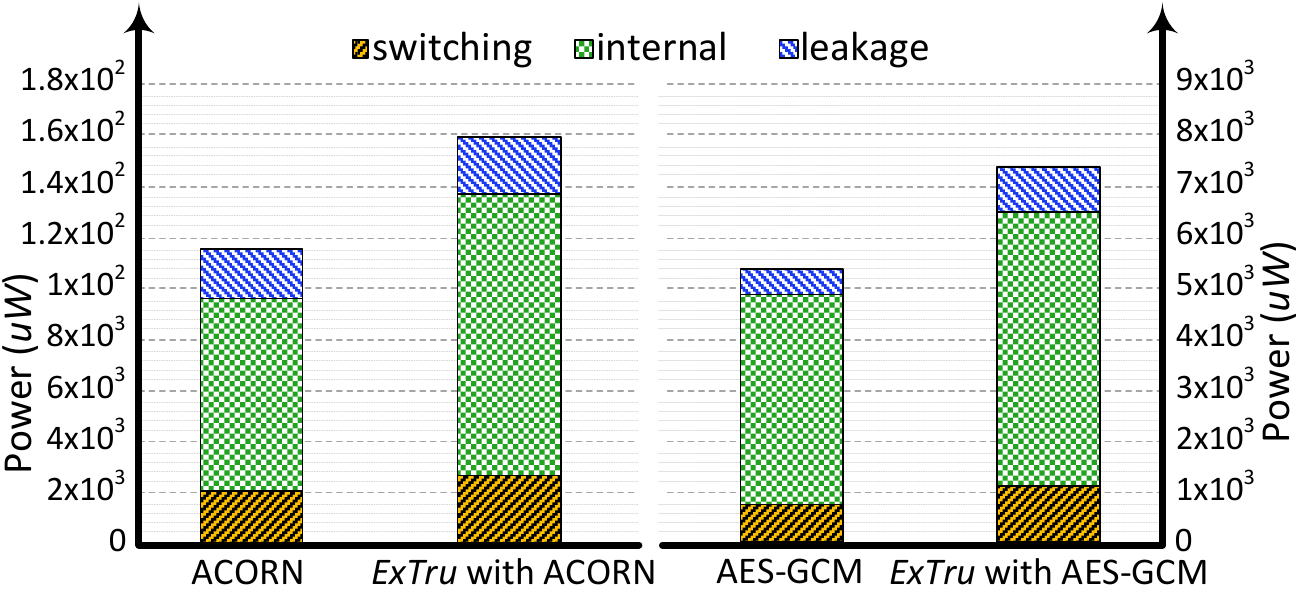}
\caption{Power Consumption of \emph{ExTru} in Comparison with Ciphers.}
\label{PL_power}
\end{figure}

Table \ref{extru_ppa} depicts resource utilization of \emph{ExTru} in each mode of using AES-GCM or ACORN. As we expected, the critical paths of \emph{ExTru} in each mode is same as that of corresponding cipher. In addition, since \emph{ExTru} consists of both CSTN and cipher, it is evident that area and power of \emph{ExTru} in each mode is approximately equal to summation of total area and total power of both sub-modules, i.e. CSTN and corresponding cipher. The active power of each design for different message sizes has been gathered using Synopsys PrimeTime PX. Fig. \ref{PL_power} demonstrates the power breakdown in each design for a 1KB message. As it is shown, the leakage powers are roughly the same. The internal power and switching power of \emph{ExTru} is almost 23\% worse. The main reason for increasing switching activity is the structure of CSTN for bit-wise permutation/toggling. Also, internal power has been increased due to merging both CSTN and cipher into one design. 

\begin{table}[t]
\footnotesize
\centering
\caption{\emph{ExTru} Resource Utilization with Different Ciphers \cite{diehl2018comparison}}
\label{extru_ppa}
\setlength\tabcolsep{3pt} 
\begin{tabular}{@{} l *9c @{}}
\toprule
\multicolumn{1}{c}{Design} & & & Power ($uW$)  & & & Area ($nm^2$)& & & Delay ($ns$) \\
\midrule
\emph{ExTru} with AES-GCM    & & &  4448.9  & & & 122457.4  & & & 2.48 \\ 
\emph{ExTru} with ACORN             & & & 1694.6    & & & 33344.7  & & & 2.3 \\ \bottomrule
\end{tabular}
\end{table}

\subsection{Energy/Performance Improvement in ExTru}

\begin{figure}[t]
\centering
\subfloat[]{{\includegraphics[width=295pt]{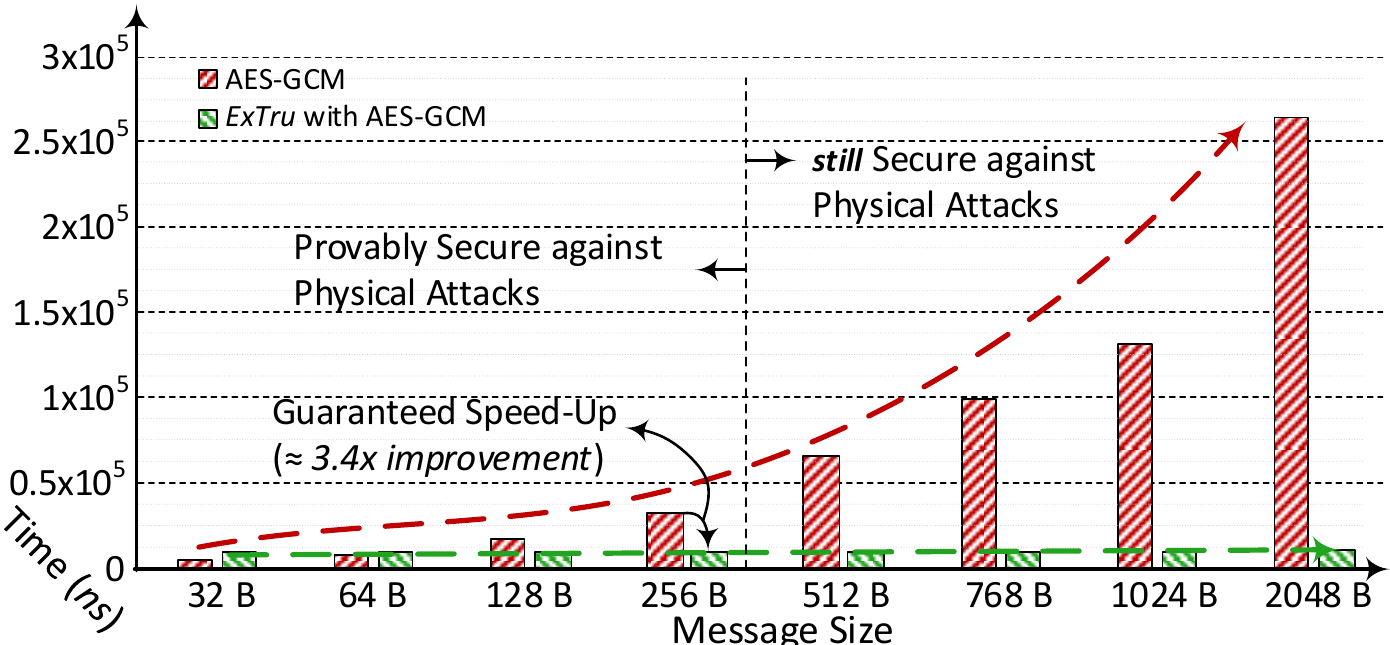} }}  \\
\subfloat[]{{\includegraphics[width=295pt]{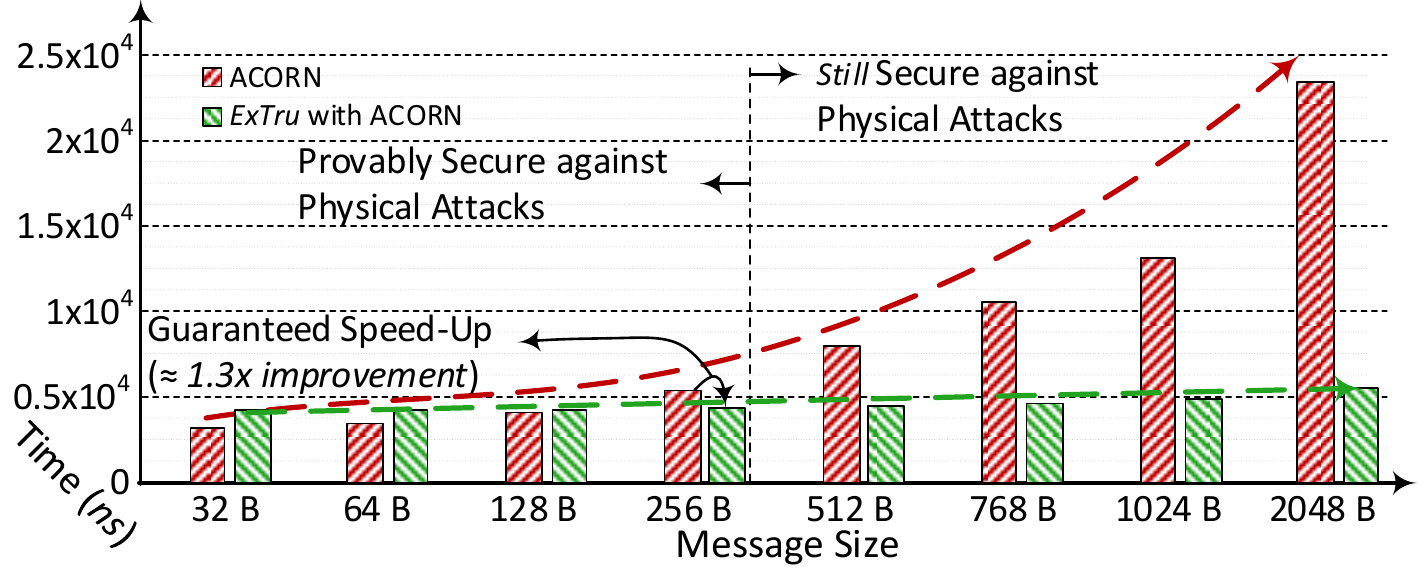} }} \newline

\caption{The Time of Preparation Data (Encryption \emph{or} Permutation/Toggling + Substitution) for Different Message Sizes (a) AES-GCM vs. \emph{ExTru} with AES-GCM (b) ACORN vs. \emph{ExTru} with ACORN.}
\label{time_exe}
\end{figure}

Although combining CSTN and cipher into \emph{ExTru} imposes area and power overhead by almost 24.5\% compared to the corresponding cipher, CSTN can generate \{permuted/toggled + substituted\} data in only one cycle which provides significant speed-up compared to especially side-channel resistant ciphers that require randomness or complex initialization. Fig. \ref{time_exe} demonstrates the time of preparing data (encryption \emph{or} permutation/toggling + substitution) for different message sizes. Increasing the size of the message, which increases the proportion of $I$ to $S$, significantly (superlinearly) increases the gap between the execution time of \emph{ExTru} compared to its corresponding cipher. As shown, since CSTN prepares each $I$ in one cycle, increasing the size of the message imposes no degradation on \emph{ExTru} performance. The main part of the execution time of \emph{ExTru} is dedicated to encrypting and sending $S$. On the other hand, all data must be encrypted before sending it while only a cipher is used. So, it increases the execution time of ciphers linearly due to encryption time. Note that based on our SAT-based evaluation, the guaranteed number of $I$ messages is 32 (Table \ref{nonblk_sat}). Since we use $LOG_{64, 4, 1}$, each $I$ is 64 bits, so 256KB ($64 \times 32$ = 2Kb = 256KB) is the safe size of sending data through CSTN. The guaranteed speed-up is $3.4\times$ and $1.3\times$ compared to AES-GCM and ACORN, respectively. 

It is evident that for small messages, \emph{ExTru} works slower than ciphers due to time overhead of sending encrypted TRN. However, \emph{ExTru} can accelerate the execution time up to $25\times$ while the message size is even 2KB. The speed-up gained by \emph{ExTru} depends on the structure of the cipher. For instance, the AES-GCM needs around 300 cycles per each plain data to be first-order side-channel resistant. However, ACORN as a stream cipher needs fewer cycles per data. So, \emph{ExTru} provides better speed-up while the cipher is not streamed/pipelined.

Table \ref{energy_cmp} depicts energy consumption for different designs, with different message sizes. Since energy is a function of time and power, it is obvious that the energy consumption in \emph{ExTru} is higher for small message sizes due to the time overhead of sending encrypted TRN. However, increasing the size of the network results in significantly less energy consumption in \emph{ExTru} compared to corresponding ciphers. As it can be seen, \emph{ExTru} reduces energy consumption by 94.5\% and 67.8\% compared to GCM and ACORN, respectively.   

\begin{table}[t]
\footnotesize
\centering
\caption{The Energy Consumption of Preparation Data (Encryption \emph{or} Permutation/Toggling + Substitution) for Different Message Sizes.}
\label{energy_cmp}
\setlength\tabcolsep{4pt} 
\begin{tabular}{@{} *9c @{}}
\toprule
\backslashbox[45pt]{Design}{Size*} & 32B & 64B & 128B & 256B & 512B & 768B & 1KB & 2KB \\ 
\toprule
ACORN                       & 17.01 & 18.69 & 22.06 & 28.79 & 42.26 & 55.73 & 69.20 & 123.1  \\
\emph{ExTru} with ACORN     & 30.66 & 30.80 & 31.09 & 31.67 & 32.82 & 33.98 & 35.13 & 39.75  \\ \midrule
AES-GCM                     & 46.28 & 93.28 & 188.2 & 379.8 & 756.6 & 1143  & 1523  & 3055  \\
\emph{ExTru} with AES-GCM   & 151.1 & 151.4 & 152.1 & 153.3 & 155.8 & 158.4 & 160.9 & 173.1 \\ 
\bottomrule
\textit{*  Message Size}
\end{tabular}
\end{table}

\begin{table}[t]
\footnotesize
\centering
\caption{Resource Utilization of \emph{ExTru} compared to corresponding ciphers in Nexys-4 DDR with Xilinx Artix 7 (XC7A100T-1CSG324).}
\label{fpga_res}
\setlength\tabcolsep{5pt} 
\begin{tabular}{@{} l *9c @{}}
\toprule
\multicolumn{1}{c}{Design} & & & LUTs & & & Registers & & & Maximum Frequency \\
\midrule
ACORN                       & & & 1090  & & & 530   & & & 178.5 MHz \\ 
\emph{ExTru} with ACORN     & & & 1609  & & & 1573  & & & 172.5 MHz \\ \midrule
AES-GCM                     & & & 3803  & & & 4418  & & & 158.3 MHz\\
\emph{ExTru} with AES-GCM   & & & 4376  & & & 5461  & & & 152.4 MHz\\ \bottomrule
\end{tabular}
\end{table}

As mentioned previously, \emph{ExTru} has been verified on both ASIC and FPGA. Table \ref{fpga_res} demonstrates the resource utilization of the proposed scheme compared to ciphers on Nexys-4 DDR with Xilinx Artix 7. The results in FPGA are approximately similar to that of ASIC. As expected, ACORN provides higher maximum frequency due to its lightweight structure. However, using more resources in high-performance AES-GCM results in better throughput even with lower frequency.

\section{Conclusion}
\label{sec:conclusion}

In this paper, we proposed \emph{ExTru} as a dynamic encrypted high speed communication, which is able to provide a level of trust using near non-blocking configurable switching and toggling network (\emph{CSTN}). \emph{ExTru} uses near non-blocking CSTN as a transceiver data. Although the configuration of CSTN will be  generated by TRNG, \emph{ExTru} changes the configuration based on a time-interval which is identified by the SAT to guarantee the security of communication. Using this dynamically encrypted mechanism mitigates energy consumption by 94.5\% and  67.8\% compared to AES-GCM (authenticated) and ACORN (stream) while security is guaranteed. In addition, \emph{ExTru} is able to provide up to $24.4\times$ and $4.3\times$ speed-up for 2KB messages in comparison with AES-GCM and ACORN, respectively.

\bibliographystyle{splncs04}
\bibliography{s-bibliography}

\end{document}